\documentclass[aps,prl,twocolumn,float]{revtex4}

\usepackage{amsmath,bm,epsfig}

  \def\Fbox#1{\vskip1ex\hbox to 8.5cm{\hfil\fboxsep0.3cm\fbox{%
    \parbox{8.0cm}{#1}}\hfil}\vskip1ex\noindent}  

\newcommand{\B}[1]{{\bm{#1}}}
\renewcommand{\sb}[1]{_{\rm{#1}}}  
\newcommand{\Sb}[1]{_{\scriptscriptstyle\rm{#1}}} 

\newcommand{\vK}{von-K\'arm\'an}
\newcommand{\Ret}{\mbox{Re}_\tau}

\begin{document}

\title{Velocity and Energy Profiles In Two- vs. Three-Dimensional Channels: \\Effects of Inverse vs. Direct Energy Cascade}

\author{Victor S. L'vov}

    \affiliation{Department of Chemical Physics, The Weizmann Institute of Science, Rehovot 76100, Israel}

\author{Itamar Procaccia}
    \affiliation{Department of Chemical Physics, The Weizmann Institute of Science, Rehovot 76100, Israel}

\author{Oleksii Rudenko}
    \affiliation{Department of Chemical Physics, The Weizmann Institute of Science, Rehovot 76100, Israel}

\begin{abstract}
In light of some recent experiments on quasi two-dimensional (2D) turbulent channel flow we provide here a model
of the ideal case, for the sake of comparison. The ideal 2D channel flow differs from its 3D counterpart
by having a second quadratic conserved variable in addition to the energy, and the latter has an inverse rather than a direct cascade.
The resulting qualitative differences in profiles of velocity $V$ and energy $K$ as a function of the distance from the wall are highlighted
and explained. The most glaring difference is that the 2D channel is much more energetic, with $K$ in wall units increasing logarithmically with the Reynolds number $\Ret$ instead of being $\Ret$-independent in 3D channels.

\end{abstract}
 \keywords{2d turbulence, 2d channel flow, mean velocity profile, algebraic model, log-law}

\maketitle

Experimental realizations of two-dimensional (2D) turbulence are not easy to come by, but there is a tradition, starting with Couder and coworkers \cite{81Cou,89CCR}, to achieve such realizations using soap-films \cite{02KG}. A number of elegant experiments, starting with Goldburg and coworkers, \cite{95KWG,01RWD,02GAK,04HWLT} on forced soap films bounded by straight wires, ignited an interest in two dimensional turbulence in a channel geometry. Indeed, a number of simulations \cite{99BGK,08KCH,08PP} and models \cite{08GG} were presented to compare with the experimental findings. While it is understood that such experiments suffer from three-dimensional (3D)
effects like film thickness fluctuations \cite{02GAK} and friction between the film and the surrounding air \cite{04HWLT}, it is clearly worthwhile to develop a reasonable theoretical model of ideal 2D channel flows to be able to gauge the degree of closeness of experiments and theory. It is quite surprising that not enough had been done in determining what are the expected velocity and energy profiles in such ideal channel flows, in parallel to the very well studied 3D channels. The aim of this Letter is to close this gap.

 We consider stationary fully developed turbulent flow of a fluid of unit density in infinitely long (in the stream-wise direction $\widehat{\mathbf{x}}$) 2D and 3D (infinitely wide in $\widehat{\mathbf{z}}$ direction) channels of width 2$L$ (in the $\widehat{\mathbf{y}}$ direction), driven by a pressure gradient $p\,'=-d p/ dx$. The velocity field is denoted as $\mathbf{U}(\mathbf{r},t) =  V(y)\widehat{\mathbf{x}}  + \mathbf{u}(\mathbf{r},t) $, where $V(y)$ is the mean velocity and $\B u(\B r,t)$ the turbulent velocity fluctuations.
In such geometries $V(y)$ and all the other mean quantities depend only on the distance from the wall $y$. We will be interested in the profiles of one-point averages, with the mean shear $ S(y) = dV/dy$, the Reynolds stress $W(y)= -\langle u_x u_y \rangle $ and
the mean turbulent kinetic energy $K(y)= \langle |\mathbf{u}^2|  \rangle /2 $ being the primary ones. In developing a model
for these profiles it is important to maintain as many exact relations as possible, and one such exact relation is the momentum balance equation \cite{Pope} which is a direct consequence of the Navier-Stokes equations:
\begin{equation}
    \label{MMB}
    \nu S(y) + W(y) = p\,'(L-y)\,, 
\end{equation}
  in which  $\nu$ is the kinematic viscosity.

  \begin{center}
    \begin{figure*}
          {\centering\includegraphics[width=7.5cm]{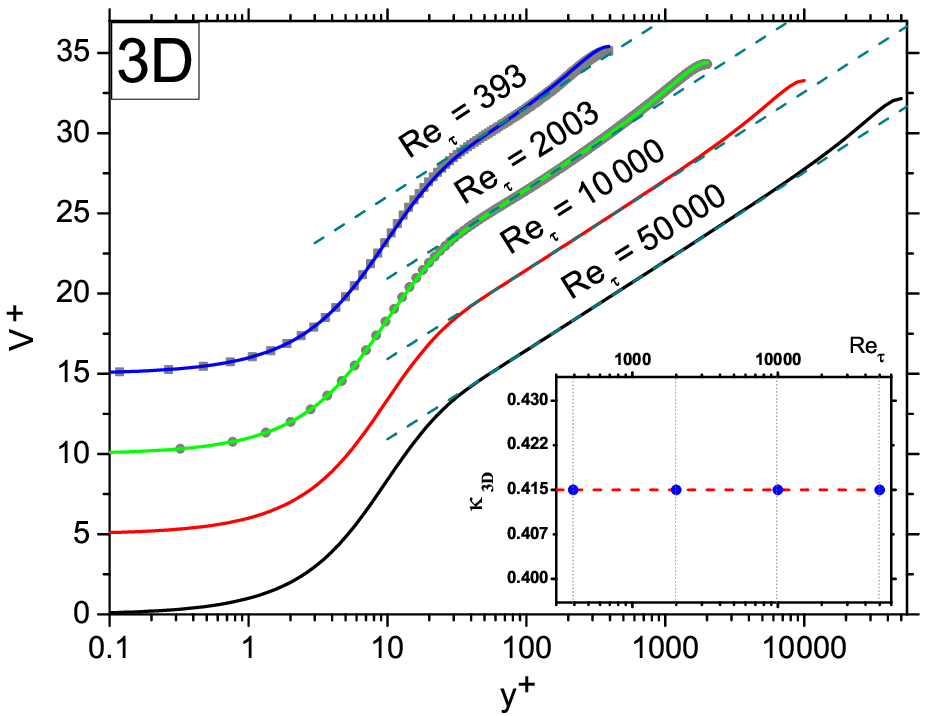}
          \includegraphics[width=7.5cm]{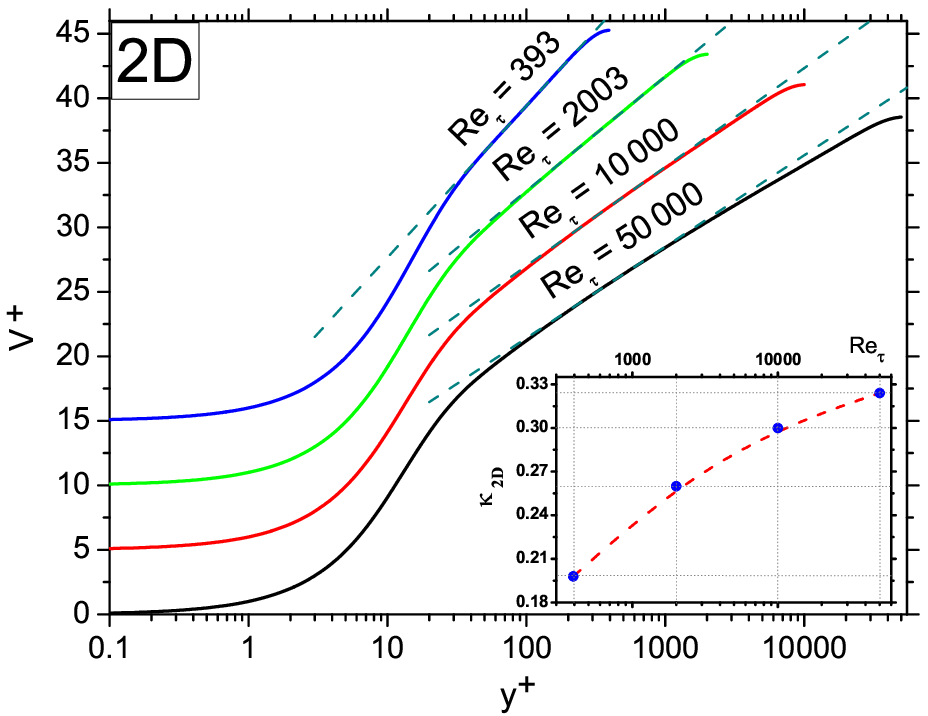}}
      \caption{Color online. Mean velocity profiles as a function of distance from the wall, in wall units, in three- and two-dimensional channels, for four values of the Reynolds number $\Ret$, shifted up by five units for clarity. The (grey) symbols in the left panel represent numerical simulations \cite{DNS}. Note that the log-law (dashed lines) in 3D with
      an invariant von-K\'arm\'an constant is increasing its range of validity whereas in 2D there is only an apparent log-law with a variable ``constant" $\kappa$ (see insets).}
      \label{velocity}
    \end{figure*}
  \end{center}

The second exact relation is the  turbulent kinetic energy balance:  $\mathcal{P}=\mathcal{D} +\varepsilon  $. Here $\mathcal{P}(y) = W S$ is the energy production, $\varepsilon(y) = \nu \left\langle (\partial_j u_i)^2 \right\rangle$ is the viscous dissipation  and $\mathcal{D}(y)=\mathcal{D}\Sb V+ \mathcal{D}\Sb T$ is the energy spatial transfer, consisting of the viscous  and turbulent contributions:
\begin{equation}
    \mathcal{D}\Sb V = -\nu \frac{d^{2} K }{d y^2}\,,\ \ 
    \mathcal{D}\Sb T = \frac{d}{d y} \Big[{{\frac{1}{2}}}{\left\langle u_y \mathbf{u}^2\right\rangle}  +\left\langle u_y \tilde{p}\, \right\rangle   \Big]\ ,
\end{equation}
where $\tilde{p}$ is the pressure fluctuation. This is as far as once can go exactly. Now we need to model $\mathcal{D}$ in terms of the one-point averages. This step is identical in 3D and 2D channels; in its simplest version it is determined by dimensional considerations:
\begin{equation}
    \label{Diff-model}
    \mathcal{D}(y) = -\frac{d}{d y}\Big[ \big( \nu\Sb T +\nu \big)\frac{d K}{d y}\Big ]\,,\ \ \nu\Sb T(y) \approx a\ \ell\, \sqrt{K}\,,
\end{equation}
where $a$ is a dimensionless constant and $\ell(y)$ is the `outer scale' whose physical meaning is the largest scale of turbulent fluctuations existing at distance $y$ from the wall. We can define $\ell(y)$ such that
$\ell(y)=y$ near the wall. Further from the wall $\ell(y)$ saturates when coming close to the channel centerline. The full $y$ dependence of $\ell(y)$ in three-dimensional channels was studied in \cite{08LPR} with the final result
$
 \ell(y) \simeq L_s\,\left\{1-\exp\left[-\lambda\,\left(1 +{\lambda}/{2} \right)\right]\right\} \ ,
$
where $L_s=0.311\,L$ and $\lambda=y(1-y/2L)/L_s$. Note that the choice of $L_s$ was made on the basis of 3D data, but we will keep the same value in 2D for lack of appropriate data. This will weakly affect the quantitative results but not at all the qualitative results below.

In the vicinity of the wall, both in two and three dimensions, we expect the flow to be differentiable, and moreover $K(y)$ should
start like $y^2$ \cite{Pope}. This allows us to compute ${\cal D}_{\rm v}$ as $-2\nu K(y)/y^2$. Since the energy production vanishes
at the wall, this forces us to estimate the energy dissipation term near the wall as
 \begin{equation}
\lim_{y\to 0}\varepsilon(y) = +2\nu K(y)/y^2\ .
 \label{dissy0} \end{equation}

To model the energy dissipation away from the wall, we will ignore (for simplicity)the tensorial structure and write (up to a factor of unity) the kinetic energy dissipation $\varepsilon \simeq \nu \int dk\, k^2\, \mathcal{K}(k)$ via the one dimensional turbulent energy spectrum $\mathcal{K}(k)$. Of course, this energy spectrum differs in two and three dimensions, resulting in a bifurcation in the further development.

\emph{3D case}. In three dimensions the direct energy cascade results in the well known K41 spectrum $\mathcal{K}\Sb{3D}(k) \simeq \varepsilon^{\,2/3}\, k^{\,-5/3}$ \cite{K41}. The turbulent kinetic energy $K(y)$ can be then estimated in the bulk as
$ \int_{1/\ell}^{\infty} dk\, \mathcal{K}\Sb{3D}(k) \sim \varepsilon^{2/3}\ell^{2/3}(y)$. The upper limit was freely extended to infinity since the integral converges in the ultraviolet. Hence, $\varepsilon \simeq b\Sb{3D}\, {K^{3/2}}/{\ell}$. Joining up the estimates near and away from the wall we can write in three dimensions
\begin{eqnarray}
  \varepsilon(y) \simeq 2\ \nu\left[{K(y)}/{\ell^{\,2}(y)}\right] +b\Sb{3D}\big[{K^{3/2}(y)}/{\ell(y)}\big]\ ,
\end{eqnarray}
where near the wall     $\ell(y)\to y$.

\begin{figure*}
 \hskip -0.4cm \includegraphics[width=0.35 \textwidth]{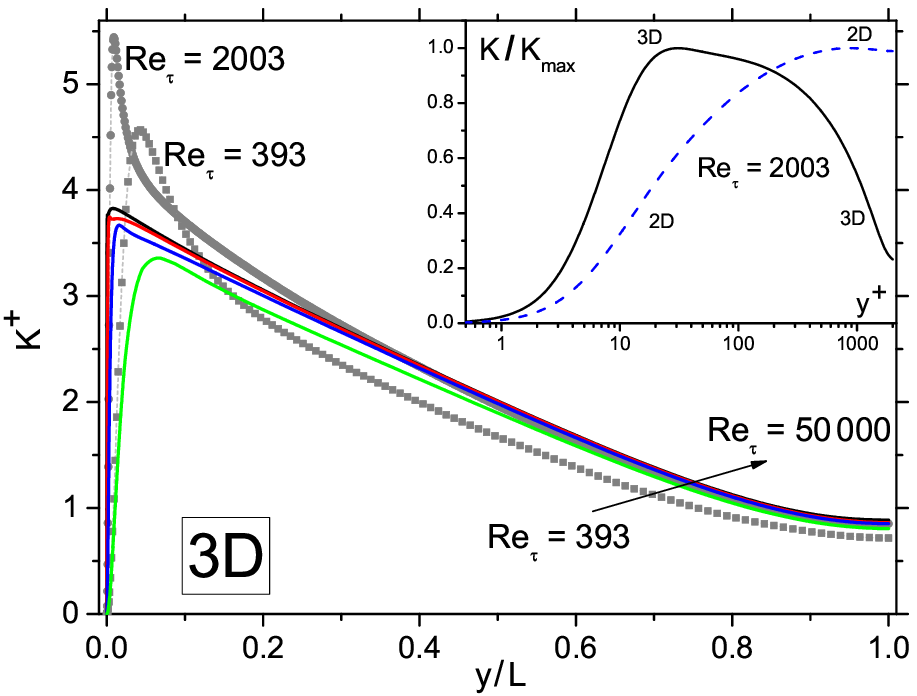}
 \hskip -0.35cm
 \includegraphics[width=0.35 \textwidth]{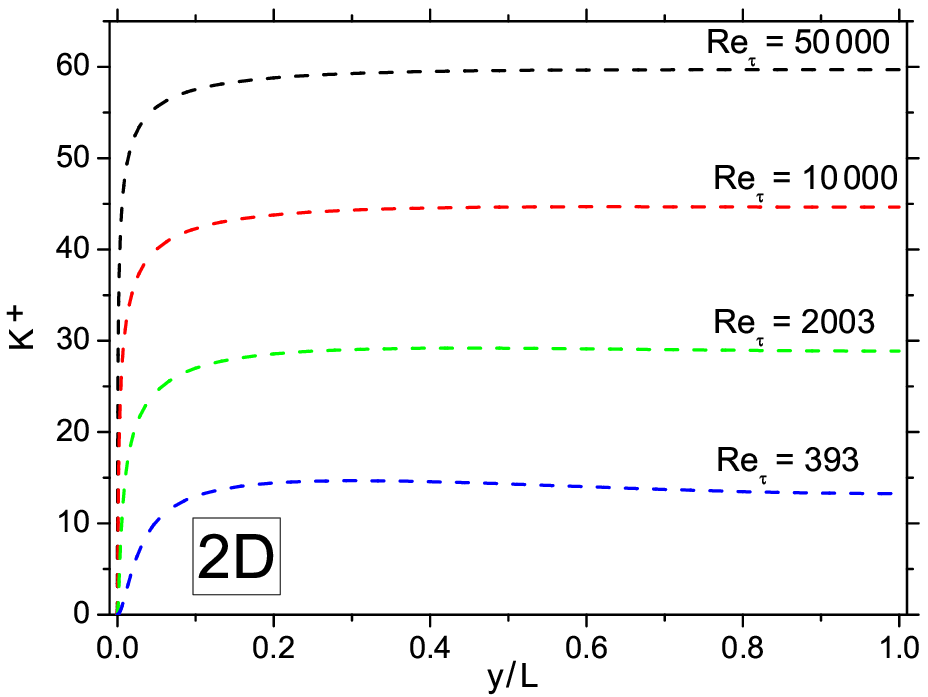}
 \hskip -0.35cm
 \includegraphics[width=0.35 \textwidth]{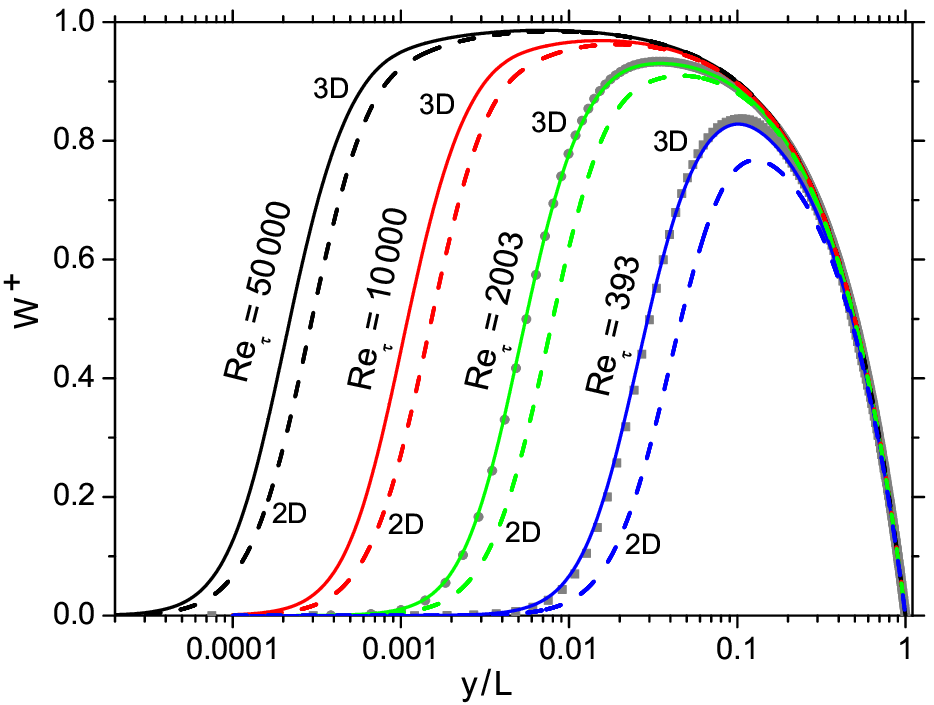}
\caption{Color online. Left and middle panels: Profiles of the turbulent kinetic energy $K^+=K/(p\,'L)$ in 3D and 2D channels respectively. Symbols in the left panel are from numerical simulations \cite{DNS}. Right panel: Reynolds shear stress $W^+=W/(p\,'L)$ as a function of $y/L$ in three (solid lines) and two-dimensions (dashed lines), respectively, for four different values of $\Ret$. The symbols are numerical simulations \cite{DNS}. The inset in the left panel compares the profiles of $K$ in three- and two-dimensions at $\Ret=2003$ normalized to their maximal values.}
 \label{energy}
 \end{figure*}

\emph{2D case}. In a two dimensional channel the inverse energy cascade does not play a role since the dominant driving is on a scale of
the order of $\ell(y)$.  Away from the walls there exists a \emph{direct enstrophy} cascade characterized by the Kraichnan \cite{Kr67} energy spectrum $\mathcal{K}\Sb{2D}(k;y) \simeq \beta^{2/3} k^{-3} \ln^{-1/3}[k \ell(y)]$,
where $\beta$ is the rate of enstrophy transfer. Therefore
\begin{eqnarray}
  \varepsilon(y) &\simeq& \nu \int_{1/\ell}^{1/\eta}\!\! dk\, k^2 \mathcal{K}\Sb{2D}(k;y) \simeq \nu \beta^{2/3} \ln^{2/3}\big [ {\ell(y)}/{\eta}\big] \nonumber\\ 
  &\simeq& \nu\, \beta^{2/3} \ln^{2/3}\big[ {\ell^{\,2}(y) \beta^{1/3}}/{\nu}\big ] \,,
  \label{eps2d}
\end{eqnarray}
where $\eta \simeq \sqrt{\nu/\beta^{1/3}\,}$ is the viscous (Kraichnan) length defined by the viscosity and  enstrophy transfer rate. To eliminate $\beta$ from these expressions we estimate $K(y)$ as
$ \int_{1/\ell}^{\infty} dk\, \mathcal{K}\Sb{2D}(k;y)\simeq \beta^{2/3} \ell^2(y)$ for $\ell(y)\gg \eta$. Using this in Eq. (\ref{eps2d}) we get
\begin{equation}
    \varepsilon(y) = b\Sb{2D}\,\nu\, {K}{\ell^{-2}} \ln^{2/3}\!\left( {\nu^{-1}}{\ell} \sqrt{K}+d\right) \ .
\end{equation}
Note that we have added a constant $d$ in the argument of the logarithm which was only correct for large values of $\ell(y)$.
With the regularizer we can go to the limit $\ell(y)\to y\to 0$ without generating a spurious divergence. Requiring now that this equation agrees with the limit~(\ref{dissy0}) we have to choose $b\Sb{2D}=2$ and $d= {e}$. In other words,
\begin{equation}
  \varepsilon(y) =  2\,\nu\big[{K(y)}/{\ell^{\,2}(y)}\big] \ln^{2/3}\!\left( {\ell(y)}\sqrt{K(y)}\big/\nu +e\right)  .
\end{equation}
The resulting energy balance in 3D and 2D is
\begin{eqnarray}
 &&\!W S +\frac{d}{d y}\left( d\ \ell \, \sqrt{K } +\nu \right)\frac{dK }{d y}\nonumber\\
               &=& \left\{\begin{array}{c c}
              \displaystyle ~~~2 \nu\,\frac{K }{\ell^{\,2} } +b\,\frac{K^{3/2} }{\ell }\ , &\quad {\rm 3D} \ , \\
              \displaystyle 2 \nu\, \frac{K }{\ell^{\,2} } \ln^{2/3}\!\left( \nu^{-1}{\ell }\sqrt{K } +\mathrm{e}\right) & \quad {\rm 2D} \ .
              \end{array}\right .
              \label{amazing}
\end{eqnarray}
Together with the momentum balance equation (\ref{MMB}) we now have two equations relating our three objects $S$, $W$ and $K$. To complete the set of equations we employ a version of the Prandtl closure $W(y)\simeq \nu_{\rm T}(y)S(y)$, which was carefully justified in \cite{08LPR} for three-dimensional channels, with the final result
\begin{equation}
    \label{W-local}
   r\Sb{W} W \approx c\, \ell \sqrt{K}\, S\,, \ \ \ r\Sb{W}(y) = \big[ 1 +\left( \ell\Sb{buf}/y\right)^6\,\big]^{1\!/6},~
\end{equation}
where $c$ and $\ell\Sb{buf}$ are constants ($\ell\Sb{buf}$ in wall units is 43 for 3D current best fits).

Having three equations for three unknown functions of $y$ we can solve them given reasonable values of the parameters. In three-dimensions we take the result of Ref.~\cite{08LPR}, for the \vK~ constant $\kappa\Sb{3D} \equiv (c^3\!/b)^{1/4}$. Experimentally in three-dimensions $\kappa\Sb{3D} \approx 0.415$, and one is left with adjusting the constants $a$ and $b$. We fix them using numerical simulation at the largest available Reynolds number, $\Ret = 2003$ \cite{DNS}, finding $a \approx 0.218$, $b \approx 0.310$. Not having independent data in two-dimensions we take there the same values for $a$, $\ell\Sb{buf}$ and $L\sb{s}$.
It was suggested in Ref.~\cite{08GG} that $\kappa\Sb{2D} \approx 0.2$ for $\Ret \approx 10^3$. While we do not agree with this reference on the existence of a power law in 2D, there is a small range of $y$ where such a law can be fitted also in our results, and we use
this number to adjust the value of $c$ to $c \approx 0.047$. This completes the choice of parameters in two an three dimensions.

The theoretical predictions for the mean velocity profiles are shown in Fig.~\ref{velocity}, where we have used the wall-coordinates
$\Ret \equiv {L\sqrt{\mathstrut p\,' L}}/{\nu}$, $y^+ \equiv {y \Ret }/{L}$,  $V^+ \equiv {V}/{\sqrt{\mathstrut p\,'L}}$. Note the good agreement between predictions and data in three-dimensions throughout the entire channel, including the viscous, buffer, log-law and wake regions. This underlines the quality of the model employed here. We note that in 3D the log-law region (shown as dashed line) increases with $\Ret$. Indeed, the present model is very similar to the one proposed in \cite{08LPR} in which the \vK~ log-law $\kappa \approx 0.415$ was obtained. The model also captures the ``wake" feature of the velocity profiles.
In contrast, the mean velocity profiles for two-dimensional channel in the right panel reveal a very limited log-law region (if any) as well as an increase in the apparent \vK~ constant upon increasing $\Ret$, see inset in Fig.~\ref{velocity} right panel.
The ``wake" feature is also absent, instead the curves bend down.

Even a larger qualitative difference is seen in the turbulent kinetic energy profiles in wall units, as seen in Fig.~\ref{energy}, left and middle panels. We first note the order of magnitude difference in the value of the kinetic energy in favor of the two-dimensional channel. Second, the 3D profiles are almost $\Ret$-independent, saturating at $\Ret\to \infty$. In contradistinction the 2D profiles increase almost linearly with $\ln(\Ret)$. The third difference is the decline of the 3D $K^+$ with $y^+$ outside the buffer layer, reaching a pronounced {\em minimum} at the centerline. In the two-dimensional case $K^+$ becomes almost $y^+$ independent. Interestingly enough, the highly different behavior of $K^+$ is not mirrored in the profiles of $W^+$ which are shown in the right panel of Fig.~\ref{energy}.

\begin{figure*}
          {\centering\includegraphics[width=0.32 \textwidth]{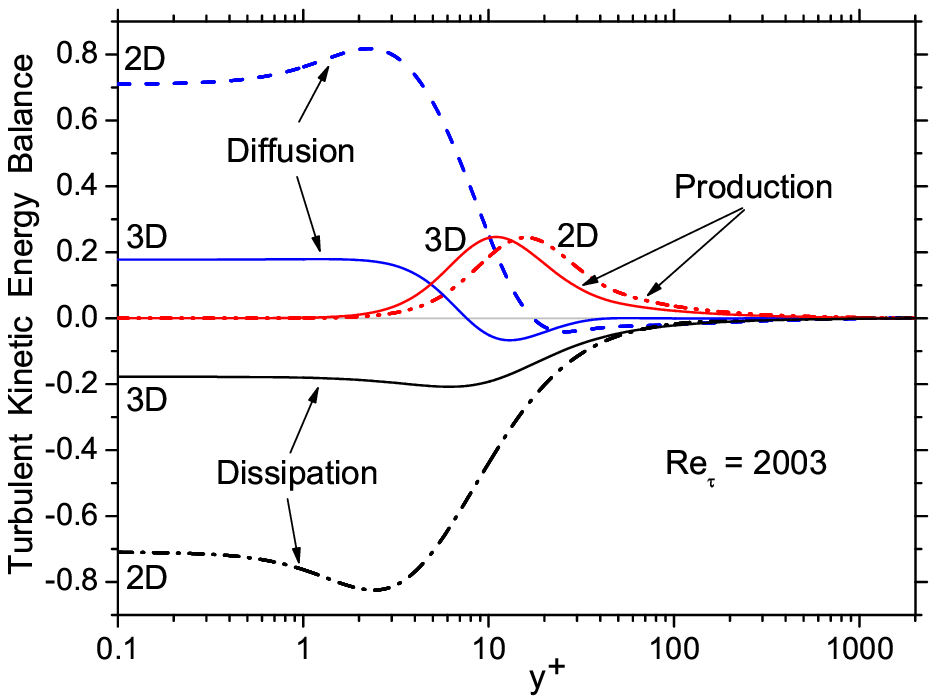}}
          ~~~~{\centering\includegraphics[width=0.32 \textwidth]{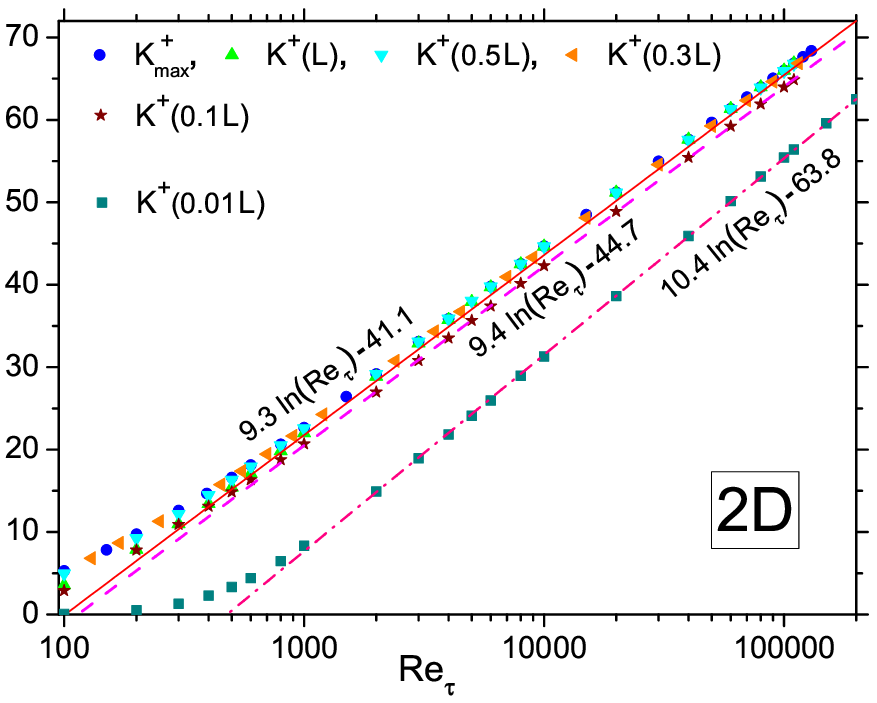}}
          \!\!\!\!{\centering\includegraphics[width=0.32 \textwidth]{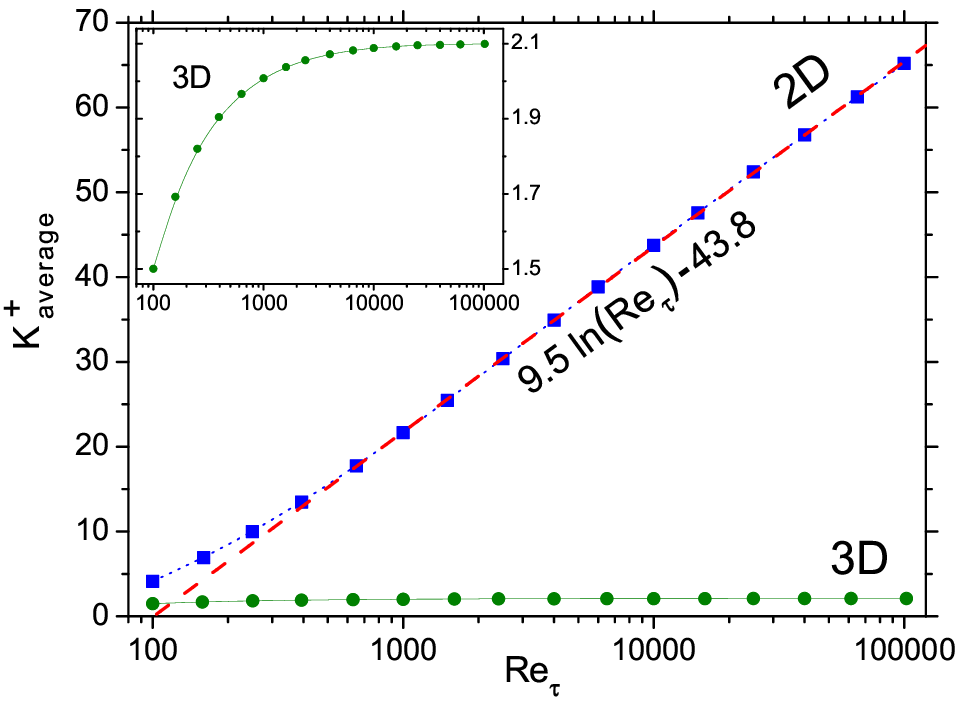}}
 \caption{Color online. Wall units. Left panel: The energy balance between production, diffusion and (minus) dissipation for 3D (solid lines) and 2D (broken lines) channels at $\Ret = 2003$.  Middle panel: The $\Ret$ dependence of typical values of the kinetic energy in 2D channels. Note the agreement with the estimate in Eq.~(\ref{rough}). Right panel: Averaged turbulent kinetic energy over the channel half-width $L$: $K\sb{average} = {L^{-1}} \int_0^L\! K(y)\, dy$. Inset: the quick saturation of $K\sb{average}^+$ in 3D channels.}
 \label{Wdiss}
 \end{figure*}

In order to rationalize all the differences presented above, we will discuss the energy balances
between production, diffusion and dissipation which are shown in the left panel of Fig.~~\ref{Wdiss}.
In order not to get mixed up between diffusion and dissipation we will consider the total (integral over the channel) energy balance from which the diffusion term disappears exactly. The integral of energy production $E\sb{\,P}\equiv \int_0^L W S dy$ is dominated by the bulk region $y> y_*=y^+_* p\,'L/\nu$ where $y^+_*\simeq 20$. In this region we can take $W\sim p\,'L$ and $S(y) \sim \sqrt{p\,'L}/(\kappa\, \nu\, y)$ where $\kappa$ is the von-K\'arm\'an constant in three-dimensions and a very weakly $\Ret$ dependent number in two-dimensions. Integrating between $y_*$ and $L$ we end up with the estimate $[(p\,'L)^{3/2}/\kappa]\ln{(L/y_*)}$ or
\begin{equation}
E\sb{\,P}\simeq\big[(p\,'L)^{3/2}/\kappa\big]\ln {(\Ret/y^+_*)} \ .
\label{Ep}
\end{equation}
On the other hand the total dissipation $E\sb{\,D}$ differs in three- and two-dimensions. In 3D we estimate the main term from the bulk of the channel as $E\sb{\,D}\approx b\int_{y_*}^L K^{3/2}(y)/y\approx \widetilde K^{3/2} \ln {(\Ret/y^+_*)}$, where $\widetilde K$ is some
typical value of $K(y)$ in the bulk. Comparing with Eq.~(\ref{Ep}) we see that indeed $\widetilde K\simeq p\,'L$ and therefore $\Ret$-independent, cf. Fig.~\ref{energy}, left panel. This is not the case in two-dimensions. According to Eq.~(\ref{amazing}) the dissipation in 2D is roughly
constant up to $y\approx \widetilde y\sim 10\,p\,'L/\nu$, whereas for $y>\widetilde y$ it decreases roughly like $y^{-2}$. Moreover, the integral over the dissipation is roughly equal in these two regions. We can therefore estimate $E\sb{\,D}\approx 4\nu \widetilde K \int_{\widetilde y}^L  dy/y^2$ where the weak logarithm is neglected. This integral is dominated by
its lower limit and therefore in two-dimensions $E\sb{\,D}\approx 4\nu \widetilde K /\widetilde y \sim \widetilde K \sqrt{p\,'L} /\widetilde y^+$. Comparing with
Eq. (\ref{Ep}) we see that
\begin{equation}
    \widetilde K^+ \approx  \left({\widetilde y^+}/{\kappa}\right) \ln {(\Ret/ y_*^+)} \sim A \ln {(\Ret)} + B\,, \label{rough}
\end{equation}
with $A$ and $B$ being constants. This explains nicely the results of Fig.~\ref{energy}, middle panel, where we see that multiplying the values of $\Ret$ by a factor of five results in equidistant curves of $K^+$ in linear scale, and this is why $\widetilde K$ can be taken as the asymptotic (almost constant) value of $K$. To support these estimates by numerics we show in Fig.~\ref{Wdiss}, middle panel, the actual functional dependence of $K^+\sb{max}$ and $K^+$ at various
position in the channel on $\Ret$. Equation (\ref{rough}) is very well supported by these data. 

Finally, we want to explain the profiles in Fig.~\ref{velocity}. The log-law in three-dimensions can easily be derived by taking the dominant terms in Eqs. (\ref{MMB}), (\ref{amazing}) and (\ref{W-local}) (in the limits $\Ret\to \infty$ and $p\,'L/\nu \ll y\ll L$);
$W=p\,'L$, $WS = b K^{3/2}/y$ and $W=c y\sqrt{K}S$. Solving these for $S$ we find $S(y)\propto 1/y$ i.e log profile. This
simple solution is lost in 2D because of the absence of the direct cascade that lead to the term $bK^{3/2}/y$. Nevertheless
we see that the role of the direct cascade in two-dimensions is mimicked by the (negative of the) diffusive term which is also $\Ret$ independent and leading to a similar estimate if we replace $dK/dy$ by $K/y$. For that reason we still have a remnant of a log-law also in 2D, but with a slope that changes with $\Ret$ due to the dependence of $\widetilde K$ on $\Ret$.

In conclusion, we show that the Reynolds stress profiles in 2D and  3D channels look similar. The velocity profiles in 3D are truly represented by a log-law, but in 2D they are only apparently varying according to a log-law, the ``constant" is changing as a function of $\Ret$, even though not vary rapidly. But the major change between two and three dimensions is in the kinetic energy profile. The two-dimensional channel is much more energetic, with the mean kinetic energy increasing like $\ln(\Ret)$. This is the main result of the loss of the direct
energy cascade. Nevertheless even without this cascade the energy dissipation exceeds its laminar value and it changes the dependence of the kinetic energy on $\Ret$ from ${\mathcal O}(\Ret)$ as in laminar flows, to ${\mathcal O}(\ln\Ret)$ which is one of the interesting predictions offered above. To make this point clear we show in the right panel of Fig. 3 the difference between the averaged kinetic energy in the channel in 2D and 3D as a function of $\Ret$. The former is roughly linear in $\ln (\Ret)$ and the the latter saturates quickly to a constant of about 2. It should be tempting to test in experiments some of these predictions, even if the ideal 2D channel model is not easy to achieve.

This work had been supported in part by Minerva Foundation, Munich, Germany. We thank Walter Goldburg and Nigel Goldenfeld for interesting communications that attracted our attention to the present issue.

\end{document}